\documentclass[showpacs,preprintnumbers,amsmath,amssymb]{revtex4}

\usepackage{amsmath,amssymb,graphicx}
\newcommand{\bea}{\begin{eqnarray}}
\newcommand{\eea}{\end{eqnarray}}

\begin{document}

\title{Axion as a Cold Dark Matter Candidate: Proof to Second order}
\author{Hyerim Noh${}^{1}$, Chan-Gyung Park${}^{2}$ and Jai-chan Hwang${}^{3,4}$}
\address{${}^{1}$Korea Astronomy and Space Science Institute, Daejon 305-348,
         Republic of Korea \\
         ${}^{2}$Division of Science Education and Institute of Fusion  Science, Chonbuk National University, Jeonju 561-756, Republic of Korea \\
         ${}^{3}$Department of Astronomy and Atmospheric Sciences, Kyungpook National University, Daegu 702-701, Republic of Korea \\
         ${}^{4}$Korea Institute for Advanced Study, Seoul 130-722, Republic of Korea}

\begin{abstract}

We {\it prove} that the axion as a coherently oscillating scalar field acts as a cold dark matter (CDM) to the second-order perturbations in all cosmological scales including the super-horizon scale. The proof is made in the axion-comoving gauge. For a canonical mass, the axion pressure term causes deviation from the CDM only on scales smaller than the Solar System size. Beyond such a small scale the equations of the axion fluid are the same as the ones of the CDM based on the CDM-comoving gauge which are exactly identical to the Newtonian equations to the second order. We also show that the axion fluid does not generate the rotational (vector-type) perturbation even to the second order. Thus, in the case of axion fluid, we have the relativistic/Newtonian correspondence to the second order, even considering the rotational perturbation. Our analysis is made in the presence of the cosmological constant, and can be easily extended to the realistic situation including other components of fluids and fields.

\end{abstract}

\noindent \pacs{14.80.Mz, 95.35.+d, 98.80.Jk, 98.80.-k \newline
Keywords: Axion, Cold dark matter, Cosmological perturbation}

\maketitle

%
%
\section{Introduction}

Axion as a coherently oscillating scalar field is one of the best known candidates for the cold dark matter (CDM) \cite{Axion-CDM}. The case was previously shown in relativistic perturbation theory to the linear order \cite{Nambu-Sasaki-1990,Ratra-1991,axion-1997,Sikivie-Yang-2009,axion-2009}.
To the linear order it can be shown in any gauge condition: in the literature the zero-shear gauge \cite{Nambu-Sasaki-1990,Sikivie-Yang-2009}, the synchronous gauge \cite{Ratra-1991}, the uniform-curvature gauge \cite{axion-1997}, and the axion-comoving gauge \cite{axion-2009} were used for the proof. However, only in the axion-comoving gauge the axion behaves as the Newtonian CDM in {\it all} cosmological scales \cite{axion-2009}. The characteristic axionic Jeans scale is merely Solar-System size for the canonical axion mass \cite{Khlopov-etal-1985,Nambu-Sasaki-1990,Sikivie-Yang-2009,axion-2009,axion-low-mass-2012}.

To the second-order perturbation the exact relativistic/Newtonian correspondence is known for the density and velocity perturbation in the comoving gauge condition for a zero-pressure fluid without rotation \cite{Noh-Hwang-2004}. The correspondence is available {\it only} in the comoving gauge; considering the density as well as the velocity perturbations this is true even to the linear order \cite{Noh-etal-2012}.

Here, based on the axion-comoving gauge we provide a general relativistic proof that axion behaves as a CDM to the second-order perturbation in all cosmologically relevant scales. Due to the exact relativistic/Newtonian correspondence for the CDM our conclusion is valid in the Newtonian context as well in all cosmological scales. As the axion fluid does not support the rotational (vector-type) perturbation, now the correspondence is valid considering the rotational perturbation as well. We set $c \equiv 1 \equiv \hbar$.

\section{Scalar field equations to the second order}

We consider scalar- and vector-type perturbations in a spatially flat Friedmann background; i.e., we ignore the transverse-tracefree perturbation of the spatial metric.
Our metric convention to the nonlinear order is \cite{Bardeen-1988} \bea
   & & d s^2 = - \left( 1 + 2 \alpha \right) d t^2
       - 2 a \left( \beta_{,\alpha} + B^{(v)}_\alpha \right)
       d t d x^\alpha
       + a^2 \left[ \left( 1 + 2 \varphi \right) \delta_{\alpha\beta}
       + 2 \gamma_{,\alpha\beta}
       + C^{(v)}_{\alpha,\beta} + C^{(v)}_{\beta,\alpha}
       \right]
       d x^\alpha d x^\beta,
\eea where we have $B^{(v)\alpha}_{\;\;\;\;\;\;\;,\alpha} \equiv 0 \equiv C^{(v)\alpha}_{\;\;\;\;\;\;\;,\alpha}$ for vector-type perturbation; the indices of perturbation variables $B^{(v)}_\alpha$ and $C^{(v)}_\alpha$ are raised and lowered by $\delta_{\alpha\beta}$ as the metric. We set $\chi \equiv a ( \beta + a \dot \gamma )$.

To the nonlinear order the scalar-, vector- and tensor-types of perturbations couple with each other in the equation level. To the second order, the nonlinear couplings occur as quadratic combinations of linear-order perturbations of all three-types of perturbation. Thus, here we ignore the presence of the coupling with the tensor-type perturbation. Later we will show that the axion does not support the vector-type perturbation even to the second order in perturbation: see Eq.\ (\ref{vector-type}).

Our convention of the energy-momentum tensor is \bea
   \widetilde T_{ab} = \widetilde \mu \widetilde u_a \widetilde u_b
       + \widetilde p \left( \widetilde g_{ab} + \widetilde u_a \widetilde u_b \right)
       + \widetilde \pi_{ab},
\eea where $\widetilde \mu$, $\widetilde p$, $\widetilde u_a$ and $\widetilde \pi_{ab}$ are the energy density, pressure, fluid four-vector, and the anisotropic stress, respectively. We introduced the perturbed fluid three-velocity as $\widetilde u_\alpha \equiv  a v_\alpha
\equiv a ( - v_{,\alpha} + v^{(v)}_\alpha )$ with $v^{(v)\alpha}_{\;\;\;\;\;\;\;,\alpha} \equiv 0$, and the aniostropic stress as $\widetilde \pi_{\alpha\beta} \equiv a^2 \Pi_{\alpha\beta}$; the indices of $v_\alpha$ and $\Pi_{\alpha\beta}$ are raised and lowered by $\delta_{\alpha\beta}$ as the metric.

In terms of the perturbed fluid quantities, to the second order we have [see Eqs.\ (54) and (84) in \cite{Hwang-Noh-2007}] \bea
   & & \widetilde T^0_0
       = - \mu - \delta \mu
       - \left( \mu + p \right) v^\alpha \left( v_\alpha + \beta_{,\alpha} + B^{(v)}_\alpha \right),
   \nonumber \\
   & & \widetilde T^0_\alpha
       = \left( \mu + p \right) \left( 1 - \alpha \right) v_\alpha + \left( \delta \mu + \delta p \right) v_\alpha + \Pi_{\alpha\beta} v^\beta,
   \nonumber \\
   & & \widetilde T^\alpha_\beta
       = \left( p + \delta p \right) \delta^\alpha_\beta + \left( \mu + p \right) \left( v^\alpha + \beta^{,\alpha} + B^{(v)\alpha} \right) v_\beta
       + \Pi^\alpha_\beta - \left( 2 \varphi \delta^\alpha_\gamma + 2 \gamma^{,\alpha}_{\;\;\;\gamma}
       + C^{(v)\alpha}_{\;\;\;\;\;\;\;,\gamma}
       + C^{(v),\alpha}_\gamma \right) \Pi_\beta^\gamma.
   \label{Tab}
\eea When we consider the axion in a fluid context we should regard the average of each individual combination of fluid quantities in the right-hand-side (RHS) of $\widetilde T^a_b$ contributing to the energy-momentum tensor.

As the spatial gauge condition we take \bea
   \gamma = 0 \equiv C^{(v)}_\alpha,
\eea thus $\beta = \chi/a$. This choice of the spatial gauge condition is important to make all remaining variables gauge-invariant even to the nonlinear order \cite{Bardeen-1988,Noh-Hwang-2004}. Furthermore, the proper relativistic/Newtonian correspondences of density and velocity perturbation to the second order are available only in this spatial gauge choice \cite{SG-2006,Noh-etal-2012}.

Later we will show that for an axion fluid the vector-type contributions vanish to the second-order perturbation: see Eq.\ (\ref{vector-type}) and below it. Thus, for an axion fluid the vector-type perturbation does not affect the scalar-type perturbation to the second order. Therefore, in the following, without losing any generality, we will consider only the scalar-type perturbation, except for presenting $(\mu + p) v^{(v)}_\alpha$ in Eq.\ (\ref{v-vector}) in order to show that it vanishes.

In our analysis we need the following equations for a general fluid [see Eqs.\ (95), (98) and (100) in \cite{Hwang-Noh-2007}] \bea
   & & \kappa - 3 H \alpha + 3 \dot \varphi
       + {\Delta \over a^2} \chi
       =
       \left( \alpha + 2 \varphi \right)
       \left( 3 \dot \varphi + {\Delta \over a^2} \chi \right)
       - {9 \over 2} H \alpha^2
       + {1 \over a^2} \chi^{,\alpha}
       \left( - \varphi + {3 \over 2} H \chi \right)_{,\alpha},
   \label{kappa-def} \\
   & & \dot \kappa
       + 2 H \kappa
       - 4 \pi G \left( \delta \mu + 3 \delta p \right)
       = - \left( 3 \dot H + {\Delta \over a^2} \right)
       \left( \alpha + {1 \over 2 a^2} \chi^{,\alpha} \chi_{,\alpha}
       \right)
       + \alpha \dot \kappa
       + {9 \over 2} \dot H \alpha^2
       + {1 \over 3} \left[ \kappa^2
       - {1 \over a^4} \left( \Delta \chi \right)^2 \right]
   \nonumber \\
   & & \qquad
       + {1 \over a^2} \left[ 2 \left( \alpha + \varphi \right)
       \Delta \alpha
       + \alpha^{,\alpha} \left( \alpha - \varphi \right)_{,\alpha}
       - \chi^{,\alpha} \kappa_{,\alpha}
       \right]
       + {1 \over a^4} \chi^{,\alpha\beta} \chi_{,\alpha\beta}
       + 8 \pi G \left( \mu + p \right) v^{,\alpha} v_{,\alpha},
   \label{kappa-eq} \\
   & & \delta \dot \mu
       + 3 H \left( \delta \mu + \delta p \right)
       - \left( \mu + p \right) \left( \kappa - 3 H \alpha
       + {\Delta \over a} v \right)
       = - {1 \over a^2} \delta \mu_{,\alpha} \chi^{,\alpha}
       + \left( \delta \mu + \delta p \right)
       \left( \kappa - 3 H \alpha \right)
       + \left( \mu + p \right) \alpha \kappa
   \nonumber \\
   & & \qquad
       + {3 \over 2} H \left( \mu + p \right)
       \left( \alpha^2 - {1 \over a^2} \chi^{,\alpha} \chi_{,\alpha} \right)
       + {1 \over a} \left( \mu + p \right) \left[
       \left( \alpha - 2 \varphi \right) \Delta v
       + \left( 2 \alpha + \varphi \right)_{,\alpha} v^{,\alpha} \right]
   \nonumber \\
   & & \qquad
       - {1 \over a^4} \left[ a^4 \left( \mu + p \right) v^{,\alpha} v_{,\alpha} \right]^{\displaystyle\cdot}
       + {1 \over a} \left[ \left( \delta \mu + \delta p \right) v^{,\alpha}
       \right]_{,\alpha}
       + {1 \over a} \left( \Pi^{\alpha\beta} v_{,\beta} \right)_{,\alpha}
       - {1 \over a^2} \Pi^{\alpha\beta}
       \chi_{,\alpha\beta},
   \label{delta-eq}
\eea
where $H \equiv \dot a/a$. These are definition of $\kappa$ (perturbed part of the trace of extrinsic curvature), the Raychaudhury equation in the normal frame (trace part of the ADM propagation equation), and the ADM energy conservation equation, respectively.

For a minimally coupled scalar field we have the equation of motion and the energy-momentum tensor given as \bea
   & & \widetilde \phi^{;c}_{\;\;\; c} = \widetilde V_{,\widetilde \phi},
   \\
   & & \widetilde T^a_b = \widetilde \phi^{,a} \widetilde \phi_{,b}
       - \left( {1 \over 2} \widetilde \phi^{,c} \widetilde \phi_{,c} + \widetilde V \right) \delta^a_b.
\eea To the background order, for the equation of motion and the fluid quantities, respectively, we have \bea
   & & \ddot \phi + 3 H \dot \phi + V_{,\phi} = 0,
   \\
   & & \mu = {1 \over 2} \dot \phi^2 + V, \quad
       p = {1 \over 2} \dot \phi^2 - V.
\eea
To the second order in perturbations, the equation of motion gives [see Eq.\ (112) in \cite{Noh-Hwang-2004}] \bea
   & & \delta \ddot \phi
       + 3 H \delta \dot \phi
       - {\Delta \over a^2} \delta \phi
       + V_{,\phi\phi} \delta \phi
       - 2 \ddot \phi \alpha
       - \dot \phi \left( \dot \alpha + 6 H \alpha
       - {\Delta \over a^2} \chi
       - 3 \dot \varphi \right)
       =
       \ddot \phi {1 \over a^2} \chi^{,\alpha} \chi_{,\alpha}
       + \dot \phi {1 \over a^2} \chi^{,\alpha}
       \left( H \chi + 2 \alpha \right)_{,\alpha}
   \nonumber \\
   & & \qquad
       + 2 \left( \alpha - \varphi \right)
       {\Delta \over a^2} \delta \phi
       - 2 V_{,\phi\phi} \alpha \delta \phi
       - 2 \dot \phi \alpha \dot \alpha
       + \dot \alpha \delta \dot \phi
       + \left( \delta \dot \phi - 2 \dot \phi \varphi \right)
       \left( \kappa - 3 H \alpha \right)
       - {2 \over a^2} \chi^{,\alpha} \delta \dot \phi_{,\alpha}
       - {1 \over 2} V_{,\phi\phi\phi} \delta \phi^2.
   \label{MSF-EOM-pert}
\eea The fluid quantities are [see Eq.\ (277) in \cite{Hwang-Noh-2007}]\bea
   & & a ( \mu + p ) v
       = \dot \phi \delta \phi
       - \Delta^{-1} \nabla^\alpha
       \left[
       \left( \delta \dot \phi - \dot \phi \alpha \right) \delta \phi_{,\alpha} \right],
   \nonumber \\
   & &
       \delta \mu
       = \dot \phi \delta \dot \phi
       - \dot \phi^2 \alpha
       + V_{,\phi} \delta \phi
       + {1 \over 2} \delta \dot \phi^2
       - {1 \over 2 a^2} \delta \phi^{,\alpha} \delta \phi_{,\alpha}
       + {1 \over 2} V_{,\phi\phi} \delta \phi^2
       - 2 \dot \phi \delta \dot \phi \alpha
       + {1 \over a^2} \dot \phi
       \chi^{,\alpha} \delta \phi_{,\alpha}
       + 2 \dot \phi^2 \alpha^2
       - {1 \over 2 a^2} \dot \phi^2 \chi^{,\alpha} \chi_{,\alpha},
   \nonumber \\
   & &
       \delta p
       = \dot \phi \delta \dot \phi
       - \dot \phi^2 \alpha
       - V_{,\phi} \delta \phi
       + {1 \over 2} \delta \dot \phi^2
       - {1 \over 2 a^2} \delta \phi^{,\alpha} \delta \phi_{,\alpha}
       - {1 \over 2} V_{,\phi\phi} \delta \phi^2
       - 2 \dot \phi \delta \dot \phi \alpha
       + {1 \over a^2} \dot \phi
       \chi^{,\alpha} \delta \phi_{,\alpha}
       + 2 \dot \phi^2 \alpha^2
       - {1 \over 2 a^2} \dot \phi^2 \chi^{,\alpha} \chi_{,\alpha},
   \nonumber \\
   & &
       \Pi^\alpha_\beta
       = 0.
   \label{fluids-MSF}
\eea
From these we can construct combinations like $(\mu + p) v^{,\alpha} v_{,\alpha}$ and $(\delta \mu + \delta p) v$ appearing in Eqs.\ (\ref{kappa-eq}) and (\ref{delta-eq}).

For $v^{(v)}_\alpha$, from Eq.\ (277) in \cite{Hwang-Noh-2007} we have \bea
   & & a \left( \mu + p \right) v^{(v)}_\alpha
       = \left( \delta \dot \phi - \dot \phi \alpha \right)
       \delta \phi_{,\alpha}
       - \nabla_\alpha \Delta^{-1} \nabla^\beta
       \left[ \left( \delta \dot \phi
       - \dot \phi \alpha \right) \delta \phi_{,\beta} \right].
   \label{v-vector}
\eea Notice that, in general, the scalar field generates vector-type perturbation to the second order. However, $(\mu + p) v_\alpha^{(v)}$ vanishes in the uniform-field gauge [setting $\delta \phi \equiv 0$ as the slicing (temporal gauge) condition] which is the same as the (field-)comoving gauge (setting $v \equiv 0$) to the second order, see Eq.\ (\ref{fluids-MSF}). In the axion case, in Eq.\ (\ref{vector-type}) we will show that, in the axion-comoving gauge [setting $\langle (\mu + p) v \rangle \equiv 0$], we have $\langle (\mu + p) v^{(v)}_\alpha \rangle = 0$ to the second-order perturbation; the time averaging symbol $\langle \; \rangle$ will be introduced soon.

\section{Axion}

\subsection{Background}

We consider the axion as a massive scalar field with $V = {1 \over 2} m^2 \phi^2$. We have \bea
   {H_0 \over m} = 2.133 \times 10^{-28} h \left( { m \over
         10^{-5} {\rm eV}} \right)^{-1},
\eea where $H_0 \equiv 100h {\rm km}{\rm sec}^{-1} {\rm Mpc}^{-1}$ is the present Hubble parameter. We strictly ignore ${H / m}$ higher order terms.

We consider the temporal average of the oscillating scalar field contributing to the fluid quantities. We have a solution \cite{Ratra-1991} \bea
   \phi (t) = a^{-3/2} \left[ \phi_{+0} \sin{(mt)}
       + \phi_{-0} \cos{(mt)} \right],
   \label{BG-phi}
\eea where $\phi_{+0}$ and $\phi_{-0}$ are constant coefficients. We take average over time scale of order $m^{-1}$ for all fluid quantities associated with the axion. We have \cite{Ratra-1991,axion-1997} \bea
   & & \mu = {1 \over 2} \langle \dot \phi^2 + m^2 \phi^2 \rangle
       = {1 \over 2} m^2 a^{-3} \left( \phi_{+0}^2 + \phi_{-0}^2 \right), \quad
       p = {1 \over 2} \langle \dot \phi^2 - m^2 \phi^2 \rangle
       = 0,
\eea
where the angular bracket indicates the time averaging. Thus, the axion evolves exactly the same as a pressureless ideal fluid \cite{Axion-CDM}. This conclusion is valid in the presence of both the spatial curvature and the cosmological constant $\Lambda$ in the background.

\subsection{Linear perturbation}

For $\delta \phi$ we take an {\it ansatz} \cite{Ratra-1991} \bea
   \delta \phi ({\bf k}, t) = \delta \phi_{+} ({\bf k}, t) \sin{(mt)}
       + \delta \phi_{-} ({\bf k}, t) \cos{(mt)}.
   \label{ansatz}
\eea As the temporal gauge condition we take the axion-comoving gauge \bea
   \langle (\mu + p) v \rangle \equiv 0.
   \label{gauge}
\eea
Each perturbation variable in this gauge can be equivalently regarded as a unique gauge-invariant combination made of the variable and $\langle (\mu + p) v \rangle$ \cite{Bardeen-1988}; this is true to the nonlinear order \cite{Noh-Hwang-2004}.
Using the gauge transformation properties in Eq.\ (252) of \cite{Noh-Hwang-2004} we have \bea
   & & \delta \mu_v \equiv \delta \mu + 3 {aH} \langle (\mu + p) v \rangle,
       \quad
       \kappa_v \equiv \kappa + {a \over \mu}
       \left( 3 \dot H + {\Delta \over a^2} \right)
       \langle (\mu + p) v \rangle,
\eea where $\delta \mu_v$ is our notation of the unique gauge-invariant combination between $\delta \mu$ and $\langle (\mu + p) v \rangle$ which becomes $\delta \mu$ in the axion-comoving gauge, etc; as all our analyses below are based on the axion-comoving gauge we ignore writing the subindex-$v$ used for indicating the gauge taken or equivalently the gauge-invariant combination.

In Eq.\ (25) of \cite{axion-2009} we have derived \bea
   \ddot \delta + 2 H \dot \delta
          - 4 \pi G \mu \delta
          = - {1 \over 4} {\Delta^2 \over m^2 a^4}
          {1 \over 1 -  {1 \over 4} {\Delta \over m^2 a^2}} \delta,
    \label{ddot-delta-eq-linear}
\eea where $\delta \equiv \delta \mu/\mu$. The term in the RHS is the characteristic pressure term arising in the axion fluid \cite{Khlopov-etal-1985,Nambu-Sasaki-1990,Sikivie-Yang-2009,axion-2009,axion-low-mass-2012}. This equation is valid in {\it all} scales including the super-horizon scale; such a general conclusion is valid in the axion-comoving gauge only. Our aim in this work is to extend this equation to the second-order perturbation: see Eq.\ (\ref{ddot-delta-eq}).

For later use we present the relations previously derived in the linear perturbation theory. Equations (16)-(22) in \cite{axion-2009} and Eqs.\ (96) and (97) in \cite{Hwang-Noh-2007} give \bea
       \delta = 2 \left( 1
       - {\Delta \over 4 m^2 a^2} \right)
       {\delta \phi_+ \over a^{-3/2} \phi_{+0}}, \quad
       \alpha = - {\delta p \over \mu}
       = {1 \over 2 m^2} {\Delta \over a^2}
       {\delta \phi_+ \over a^{-3/2} \phi_{+0}}, \quad
       \kappa = \dot \delta, \quad
       \varphi = - H {a^2 \over \Delta} \left( {3 \over 2} H \delta
       + \kappa \right), \quad
       \chi = - {a^2 \over \Delta} \kappa.
   \nonumber \\
   \label{linear-relation}
\eea

\subsection{Second-order perturbation}

We take the same {\it ansatz} in Eq.\ (\ref{ansatz}), and take the axion-comoving gauge in Eq.\ (\ref{gauge}) to the second order. Equation (\ref{fluids-MSF}) gives \bea
   & &
       \langle a ( \mu + p ) v \rangle
       = {1 \over 2} a^{-3/2} m \Big\{
       \phi_{+0} \delta \phi_-
       - \phi_{-0} \delta \phi_+
   \nonumber \\
   & & \qquad
       + \Delta^{-1} \nabla^\alpha
       \left[
       - a^{3/2} \left( \delta \phi_+ \delta \phi_{-,\alpha}
       - \delta \phi_- \delta \phi_{+,\alpha} \right)
       + \left( \phi_{+0} \delta \phi_{-, \alpha}
       - \phi_{-0} \delta \phi_{+, \alpha} \right) \alpha \right]
       \Big\}.
\eea Thus the axion-comoving gauge in Eq.\ (\ref{gauge}) gives \bea
   {\delta \phi_+ \over \phi_{+0}}
       = {\delta \phi_- \over \phi_{-0}},
\eea which is valid to the second order.

Using these we can show that Eq.\ (\ref{v-vector}) becomes \bea
   \langle \left( \mu + p \right) v^{(v)}_\alpha \rangle = 0,
   \label{vector-type}
\eea to the second order. As we have \cite{Bardeen-1980} \bea
    & & {\Delta \over a^2} \Psi^{(v)}_\alpha
        + 8 \pi G \left( \mu + p \right) v^{(v)}_\alpha
        = {\rm nonlinear \; terms},
\eea with $\Psi^{(v)}_\alpha \equiv B^{(v)}_\alpha + a \dot C^{(v)}_\alpha$, the accompanying vector-type metric perturbation also vanishes at least to the linear order. This implies that under our gauge condition the axion fluid does not support $v^{(v)}_\alpha$ to the second order, and as a consequence the vector-type perturbation does not feedback the scalar-type perturbation of the axion fluid. This has an important implication that in the case of axion fluid we have the relativistic/Newtonian correspondence valid to the second order now {\it including} the vector-type perturbation. In the ordinary zero-pressure fluid, the relativistic/Newtonian correspondence to the second order was proved by ignoring the vector- and tensor-type perturbation and the background curvature \cite{Noh-Hwang-2004,Hwang-Noh-2005-second}. In the presence of vector-type perturbation, in general, we have pure general relativistic correction terms appearing in the second order as presented in Section IX of \cite{Hwang-Noh-2007}.

Equation (\ref{fluids-MSF}) gives \bea
   & &
       {\delta p \over \mu}
       =
       - \alpha
       - {1 \over 2 a^2} \chi^{,\alpha} \chi_{,\alpha}
       - \alpha \delta + \alpha^2
       - {1 \over 2 m^2 a^2}
       {\delta \phi_+^{\;\; ,\alpha} \delta \phi_{+,\alpha}
       \over a^{-3} \phi_{+0}^2}
       = \delta
       - 2 {\delta \phi_+ \over a^{-3/2} \phi_{+0}}
       - {\delta \phi_+^2 \over a^{-3} \phi_{+0}^2}.
   \label{delta-p-1}
\eea Equations (\ref{kappa-eq}) and (\ref{delta-eq}) give \bea
   & & \dot \kappa
       + 2 H \kappa
       - 4 \pi G \mu \delta
       + {1 \over a^2} \kappa_{,\alpha} \chi^{,\alpha}
       - {1 \over a^4} \chi^{,\alpha\beta}
       \chi_{,\alpha\beta}
       = - {\Delta \over a^2} \left( \alpha
       + {1 \over 2 a^2} \chi^{,\alpha} \chi_{,\alpha} \right)
       - H \alpha \left( 2 \kappa + 3 H \delta \right)
       - {9 \over 4} H^2 \alpha^2
   \nonumber \\
   & & \qquad
       + \left( \alpha + 2 \varphi \right) {\Delta \over a^2} \alpha
       + {1 \over a^2} \alpha^{,\alpha}
       \left( \alpha - \varphi \right)_{,\alpha}
       + {3 H^2 \over 4 m^2 a^2}
       {\delta \phi_+^{\;\; ,\alpha} \delta \phi_{+,\alpha}
       \over a^{-3} \phi_{+0}^2},
   \label{dot-kappa-eq} \\
   & & \dot \delta
       - \kappa
       - \kappa \delta
       + {1 \over a^2} \delta_{,\alpha} \chi^{,\alpha}
       =
       {3 \over 2} H \alpha^2
       - \left[\!\!\left[
       {1 \over m^2} {1 \over a^2} {\delta \phi_+^{\;\;,\alpha} \over a^{-3/2} \phi_{+0}}
       \left[ \left( {\delta \phi_{+,\alpha} \over a^{-3/2} \phi_{+0}} \right)^{\displaystyle\cdot}
       - H {\delta \phi_{+,\alpha} \over a^{-3/2} \phi_{+0}} \right]
       + {\cal O} ({\rm same \; order})
       \right]\!\!\right].
   \label{dot-delta-eq}
\eea The middle term in the RHS of Eq.\ (\ref{dot-delta-eq}), coming from $\langle (\mu + p) v^{,\alpha} v_{,\alpha} \rangle$ term in Eq.\ (\ref{delta-eq}), is ${H \over m}$-order smaller than the leading order term of $\langle [(\delta \mu + \delta p) v^{,\alpha} ]_{|\alpha} \rangle$ in Eq.\ (\ref{delta-eq}) which vanishes; in our method, as we strictly ignore ${H \over m}$-order correction terms, we cannot quantitatively estimate the next order term, thus expressed as ${\cal O} ({\rm same \; order})$ in Eq.\ (\ref{dot-delta-eq}). Thus, the whole terms in the parenthesis $[\![ \; ]\!]$ should be ignored.

For a zero-pressure fluid the RHSs of Eqs.\ (\ref{dot-kappa-eq}) and (\ref{dot-delta-eq}) vanish \cite{Noh-Hwang-2004,Hwang-Noh-2005-second}.
Thus, apparently the RHSs of Eqs.\ (\ref{dot-kappa-eq}) and (\ref{dot-delta-eq}) are new contributions of the axion fluid compared with the CDM as a zero-pressure fluid. In the CDM case we have equations for $\delta$ and $\kappa$ in the CDM-comoving gauge which exactly coincide with the zero-pressure Newtonian equations for perturbed density and velocity even to the second-order perturbations \cite{Noh-Hwang-2004,Hwang-Noh-2005-second}.  By determining $\alpha$ to the second order, and the other variables to the linear order, in terms of $\delta$ and $\kappa$ we can derive the closed-form equations for $\delta$ and $\kappa$.

The variable $\alpha$ to the second order can be determined from the equation of motion. Equation (\ref{MSF-EOM-pert}) gives
\bea
   & & - m \left( 2 \delta \dot \phi_-
       + 3 H \delta \phi_- \right)
       - {\Delta \over a^2} \delta \phi_+
       = - 2 m^2 a^{-3/2} \phi_{+0} \alpha
       - m a^{-3/2} \phi_{-0} \left( \dot \alpha
       + 3 H \alpha + \kappa \right)
       - m \delta \phi_- \left( \dot \alpha - 3 H \alpha + \kappa
       \right)
   \nonumber \\
   & & \qquad
       + {2 \over a^2} m \chi^{,\alpha} \delta \phi_{-,\alpha}
       + 2 \left( \alpha - \varphi \right) {\Delta \over a^2} \delta
       \phi_+
       - 2 m^2 \alpha \delta \phi_+
       - m^2 a^{-3/2} \phi_{+0} {1 \over a^2} \chi^{,\alpha} \chi_{,\alpha}
   \nonumber \\
   & & \qquad
       - m a^{-3/2} \phi_{-0} \left[
       \alpha \kappa
       - 2 \alpha \dot \alpha
       + {3 \over 2} H \alpha^2
       + {1 \over a^2} \chi^{,\alpha} \left(
       - {1 \over 2} H \chi + 2 \alpha + \varphi \right)_{,\alpha}
       \right],
   \\
   & & m \left( 2 \delta \dot \phi_+
       + 3 H \delta \phi_+ \right)
       - {\Delta \over a^2} \delta \phi_-
       = - 2 m^2 a^{-3/2} \phi_{-0} \alpha
       + m a^{-3/2} \phi_{+0} \left( \dot \alpha
       + 3 H \alpha + \kappa \right)
       + m \delta \phi_+ \left( \dot \alpha - 3 H \alpha + \kappa
       \right)
   \nonumber \\
   & & \qquad
       - {2 \over a^2} m \chi^{,\alpha} \delta \phi_{+,\alpha}
       + 2 \left( \alpha - \varphi \right) {\Delta \over a^2} \delta
       \phi_-
       - 2 m^2 \alpha \delta \phi_-
       - m^2 a^{-3/2} \phi_{-0} {1 \over a^2} \chi^{,\alpha} \chi_{,\alpha}
   \nonumber \\
   & & \qquad
       + m a^{-3/2} \phi_{+0} \left[
       \alpha \kappa
       - 2 \alpha \dot \alpha
       + {3 \over 2} H \alpha^2
       + {1 \over a^2} \chi^{,\alpha} \left(
       - {1 \over 2} H \chi + 2 \alpha + \varphi \right)_{,\alpha} \right].
\eea From these we have \bea
   & & \alpha
       + {1 \over 2 a^2} \chi^{,\alpha} \chi_{,\alpha}
       = {1 \over 2 m^2}
       \left( 1 + 2 \alpha - 2 \varphi \right)
       {\Delta \over a^2}
       {\delta \phi_+ \over a^{-3/2} \phi_{+0}}
       - \alpha
       {\delta \phi_+ \over a^{-3/2} \phi_{+0}}.
   \label{alpha}
\eea
From Eqs.\ (\ref{delta-p-1}) and (\ref{alpha}) we can express $\alpha$ in terms of $\delta$ as
\bea
   & &
       \alpha + {1 \over 2 a^2} \chi^{,\alpha} \chi_{,\alpha}
       = {1 \over 1 - {\Delta \over 4 m^2 a^2}}
       \Bigg\{ {\Delta \over 4 m^2 a^2} \delta
       - {\delta \phi_+^2 \over a^{-3} \phi_{+0}^2}
       + {1 \over 2 m^2 a^2} {\delta \phi_+^{\;\;,\alpha}
       \delta \phi_{+,\alpha} \over a^{-3} \phi_{+0}^2}
       - {\Delta \over 4 m^2 a^2} \alpha^2
   \nonumber \\
   & & \qquad
       + \left[ \alpha + \left( {1 \over 2} \alpha - \varphi \right)
       {\Delta \over m^2 a^2} \right]
       {\delta \phi_+ \over a^{-3/2} \phi_{+0}} \Bigg\}
       + {\delta \phi_+^2 \over a^{-3} \phi_{+0}^2}
       - \alpha \delta
       - {1 \over 2 m^2 a^2} {\delta \phi_+^{\;\;,\alpha}
       \delta \phi_{+,\alpha} \over a^{-3} \phi_{+0}^2}.
   \label{a-second}
\eea For other terms in Eqs.\ (\ref{dot-kappa-eq}) and (\ref{dot-delta-eq}) we can use relations to the linear order presented in Eq.\ (\ref{linear-relation}).

\section{Axion as a CDM}

To the second order we identify the relative density perturbation $\delta$ and the velocity perturbation ${\bf u}$ as \cite{Noh-Hwang-2004} \bea
   \delta_v = \delta, \quad
         \kappa_v \equiv - {1 \over a} \nabla \cdot {\bf u}.
\eea Using Eq.\ (\ref{linear-relation}) and (\ref{a-second}), Eqs.\ (\ref{dot-kappa-eq}) and (\ref{dot-delta-eq}) provide the closed-form equations for $\delta$ and ${\bf u}$. By keeping only leading ${\Delta \over m^2 a^2}$-order correction terms, we have \bea
   & &
       {1 \over a} \nabla \cdot \left( \dot {\bf u} + H {\bf u}
       \right)
       + 4 \pi G \mu \delta
       + {1 \over a^2} \nabla \cdot \left( {\bf u} \cdot \nabla {\bf
       u} \right)
       =
       {\Delta^2 \over 4 m^2 a^4} {1 \over 1 - {\Delta \over 4 m^2 a^2}} \Bigg[ \delta
       - {1 \over 4} \left( {1 \over 1 - {\Delta \over 4 m^2 a^2}} \delta
       \right)^2 \Bigg]
   \nonumber \\
   & &
       \qquad
       + {3 H^2 \over 4 m^2} \left[\!\!\left[
       {7 \over 4 a^2} \delta^{,\alpha} \delta_{,\alpha}
       + 2 {\Delta \over a^2}
       \left[ {a^2 \over \Delta}
       \left( \delta - {2 \over 3} {1 \over aH} \nabla \cdot {\bf u} \right)
       {\Delta \over a^2} \delta \right]
       - {5 \over 2 a^2}
       \left[ {a^2 \over \Delta} \left( \delta - {2 \over 3} {1 \over aH} \nabla \cdot {\bf u} \right) \right]^{,\alpha}
       \left( {\Delta \over a^2} \delta \right)_{,\alpha} \right]\!\!\right],
   \label{Euler-eq-0} \\
   & &
       \dot \delta + {1 \over a} \nabla \cdot {\bf u}
       + {1 \over a} \nabla \cdot \left( \delta {\bf u} \right)
       =
       \left[\!\!\left[ {H \over 4 m^2 a^2} \left( \nabla \delta \right) \cdot
       \nabla \left( \delta + {1 \over aH} \nabla \cdot {\bf u}
       \right)
       + {\cal O} ({\rm same \; order})
       \right]\!\!\right].
   \label{Continuity-eq-0}
\eea Here the terms in RHSs are of
${\Delta^2 \over H^2 m^2 a^4}$-order [the leading part of the first term in Eq.\ (\ref{Euler-eq-0})] and ${\Delta \over m^2
a^2}$-order (including terms in parentheses $[\![ \; ]\!]$) smaller than the terms in the left-hand-sides (LHSs).

Based on a comment below Eq.\ (\ref{dot-delta-eq}), terms in the RHS of Eq.\ (\ref{Continuity-eq-0}), i.e., the ${\Delta \over m^2 a^2}$-order correction terms, should be ignored. Thus we should ignore the same order terms [terms in the parentheses $[\![ \; ]\!]$ and the part of nonlinear terms in the first term] in Eq.\ (\ref{Euler-eq-0}) as well. By keeping ${\Delta^2 \over H^2 m^2 a^4}$-order correction terms only we have \bea
   & & {1 \over a} \nabla \cdot \left( \dot {\bf u} + H {\bf u}
       \right)
       + 4 \pi G \mu \delta
       + {1 \over a^2} \nabla \cdot \left( {\bf u} \cdot \nabla {\bf
       u} \right)
       =
       {\Delta^2 \over 4 m^2 a^4} \left( \delta
       - {1 \over 4} \delta^2 \right),
   \label{Euler-eq} \\
   & & \dot \delta + {1 \over a} \nabla \cdot {\bf u}
       + {1 \over a} \nabla \cdot \left( \delta {\bf u} \right)
       = 0.
   \label{Continuity-eq}
\eea Except for the terms in the RHS of Eq.\ (\ref{Euler-eq}), Eqs.\ (\ref{Euler-eq}) and (\ref{Continuity-eq}) are the well known Euler and continuity equations, respectively, in Newtonian theory \cite{Peebles-1980}. The Newtonian equations are valid in the fully nonlinear situation in Newton's theory, whereas exactly the same equations are valid only to the second-order perturbations in Einstein's theory \cite{Noh-Hwang-2004}. We call it the exact relativistic/Newtonian correspondence (for both $\delta$ and ${\bf u}$) to the second-order perturbations \cite{Noh-Hwang-2004,Hwang-Noh-2005-second}; in Einstein's gravity the pure relativistic correction terms start appearing from the third order \cite{Hwang-Noh-2005-third}. By combining the above equations we have \bea
   & &
       \ddot \delta + 2 H \dot \delta
       - 4 \pi G \mu \delta
       + {1 \over a^2} \left[ a
       \nabla \cdot \left( \delta {\bf u} \right) \right]^{\displaystyle\cdot}
       - {1 \over a^2} \nabla \cdot \left( {\bf u} \cdot \nabla {\bf u} \right)
       = - {\Delta^2 \over 4 m^2 a^4} \left( \delta
       - {1 \over 4} \delta^2 \right).
   \label{ddot-delta-eq}
\eea
To the linear order we recover Eq.\ (\ref{ddot-delta-eq-linear}) in the same limit; we note that, contrary to the second-order case, the ${\Delta \over m^2 a^2}$-order correction term in the RHS of Eq.\ (\ref{ddot-delta-eq-linear}) is valid in the linear-order perturbation. Terms in RHS are the only characteristic difference caused by axion compared with CDM. Compared with terms in the LHS we have the instability scale \cite{Khlopov-etal-1985,Nambu-Sasaki-1990,Sikivie-Yang-2009,axion-2009,axion-low-mass-2012} \bea
   \lambda_J = {2 \pi a \over k_J}
       = \left( {\pi^3 \over G \mu m^2} \right)^{1/4} = 5.4 \times 10^{14} {\rm cm} h^{-1/2} \left( {m \over 10^{-5} {\rm eV}} \right)^{-1/2},
\eea with $\Delta = - k^2$.
Equation (\ref{ddot-delta-eq}) shows that the same criteria applies to the second order. The axion pressure term becomes important on scale smaller than the Solar System size for $m \sim 10^{-5} {\rm eV}$, thus completely negligible in all cosmological scales. This axionic Jeans scale increases as we decrease the axion mass, and for $m < 10^{-22}{\rm eV}$ its effect becomes important in the large scale structure, see \cite{axion-low-mass-2012}.

\section{Discussion}

In this work we have studied the axion as a fluid to the second-order perturbation in the context of Einstein's gravity. We have taken the axion-comoving gauge. The new results valid to the second-order perturbation are the following.

(i) The axion behaves as a zero-pressure fluid in all cosmologically relevant scales including the super-horizon scale. Thus, we have proved that the axion behaves as the CDM even to the second order in the perturbation.

(ii) For the canonical axion mass $m \sim 10^{-5} {\rm eV}$ the axionic Jeans scale is about the Solar-System scale, thus completely negligible in the cosmological context.

(iii) The axion fluid does not support the rotational (vector-type) perturbation. As a consequence, beyond the axionic Jeans scale the axion fluid shows the exact correspondence with the Newtonian mass and momentum conservation equations even considering the rotational perturbation. Meanwhile, the relativistic/Newtonian correspondence in the ordinary fluid was proved to the second-order perturbation {\it assuming} irrotational zero-pressure ideal fluid \cite{Noh-Hwang-2004,Hwang-Noh-2005-second}; further assumptions made are ignoring the tensor-type perturbation and the background curvature, and single component fluid. The general cases without taking all these assumptions were studied in \cite{Hwang-Noh-2007}. The vector-type perturbation, in general, introduces pure general relativistic effects to the second-order perturbation as presented in Section IX of \cite{Hwang-Noh-2007}.

Extending our study to the third order perturbation is a trivial, though tedious, exercise. In the context of a zero-pressure irrotational fluid we have shown that the pure general relativistic effects start appearing in the third order perturbation, and its effect on the matter power-spectrum is completely negligible in all cosmological scales including the super-horizon scale \cite{Hwang-Noh-2005-third}. We anticipate the same result is valid for axion fluid in all cosmological scales, and its concrete study is left for a future exercise.

Our analysis is made in the presence of the cosmological constant, and it is a trivial exercise to extend our analysis to a realistic situation in the presence of other components of fluids and fields.

%
%

%
%
\section*{Acknowledgments:}

J.H.\ was supported by KRF Grant funded by the Korean Government
(KRF-2008-341-C00022).
C.G.P. was supported by Basic Science Research Program through the National Research Foundation of Korea (NRF) funded by the Ministry of Science, ICT and Future Planning (No.\ 2013R1A1A1011107).
H.N.\ was supported by National Research Foundation of Korea funded by the Korean Government (No.\ 2012R1A1A2038497).

%
%

\end{document}